\documentclass[acmlarge]{acmart}

\setcopyright{none}
\renewcommand\footnotetextcopyrightpermission[1]{}
\usepackage{array}
\newcolumntype{L}[1]{>{\raggedright\let\newline\\\arraybackslash\hspace{0pt}}m{#1}}
\newcolumntype{C}[1]{>{\centering\let\newline\\\arraybackslash\hspace{0pt}}m{#1}}
\newcolumntype{R}[1]{>{\raggedleft\let\newline\\\arraybackslash\hspace{0pt}}m{#1}}

\AtBeginDocument{%
  }

\usepackage[normalem]{ulem}

\begin{document}
\title{Disentangling Hype from Practicality: On Realistically Achieving Quantum Advantage}

\author{Torsten Hoefler}
\affiliation{%
	\institution{Microsoft Corporation}
	\streetaddress{One Microsoft Way}
	\city{Redmond}
	\state{Washington}
	\country{USA}
	\postcode{98052}
}
\affiliation{%
	\institution{ETH Zurich}
	\streetaddress{Universitaestsstrasse 6}
	\city{Zurich}
	\state{Zurich}
	\country{Switzerland}
	\postcode{8092}
}
\email{v-torhoefler@microsoft.com}
\author{Thomas H\"aner}
\authornote{This work was done prior to T.H. joining AWS}
\email{haenerthomas@gmail.com}
\author{Matthias Troyer}
\email{mtroyer@microsoft.com}
\affiliation{%
  \institution{Microsoft Corporation}
  \streetaddress{One Microsoft Way}
  \city{Redmond}
  \state{Washington}
  \country{USA}
  \postcode{98052}
}
%
%
%
%
%
%
%

\renewcommand{\shortauthors}{Hoefler et al.}

\begin{abstract}
Quantum computers offer a new paradigm of computing with the potential to vastly outperform any imagineable classical computer. 
This has caused a gold rush towards new quantum algorithms and hardware. 
In light of the growing expectations and hype surrounding quantum computing we ask the question which are the promising applications to realize quantum advantage. 
We argue that small data problems and quantum algorithms with super-quadratic speedups are essential to make quantum computers useful in practice. 
With these guidelines one can separate promising applications for quantum computing from those where classical solutions should be pursued. 
While most of the proposed quantum algorithms and applications do not achieve the necessary speedups to be considered practical, we already see a huge potential in material science and chemistry. 
We expect further applications to be developed based on our guidelines. 

\end{abstract}

%


\maketitle

\thispagestyle{plain}
\pagestyle{plain}

Operating on fundamentally different principles than conventional computers, quantum computers promise to solve a variety of important problems that seemed forever intractable on classical computers. Leveraging the quantum foundations of nature, the time to solve certain problems on quantum computers grows more slowly with the size of the problem than on classical computers---this is called \emph{quantum speedup}. Going beyond quantum supremacy~\cite{arute19}, which was the demonstration of a quantum computer outperforming a classical one for an artificial problem, an important question is finding meaningful applications (of academic or commercial interest) that can realistically be solved faster on a quantum computer than on a classical one. We call this a practical quantum advantage, or \emph{quantum practicality} for short.

\begin{figure}[h]
	\includegraphics[width=10cm]{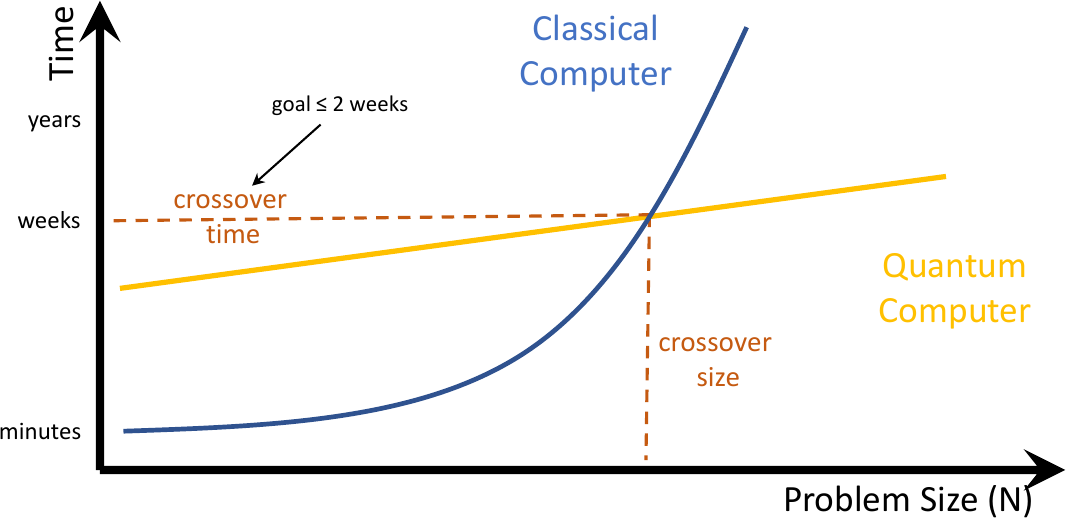}
	\caption{\textbf{Quantum speedup}: The time needed to solve certain problems with quantum algorithms increases more slowly than that of any known classical algorithm as the problem size N increases. To be practical, however, we need more than an asymptotic speedup: the crossover time where quantum advantage gets realized needs to be reasonably short and the crossover problem size not too large. {(For illustration, the time axis is scaled such that the quantum algorithm is a straight line.)}}
\end{figure}

There is a maze of hard problems that have been suggested to profit from quantum acceleration: from cryptanalysis, chemistry and materials science, to optimization, big data, machine learning, database search, drug design and protein folding, fluid dynamics and weather prediction. But which of these applications realistically offer a potential quantum advantage in practice? 
For this, we cannot only rely on asymptotic speedups but must consider the constants involved. 
Being optimistic in our outlook for quantum computers, we will identify clear guidelines for quantum practicality  and use them to classify which of the many proposed applications for quantum computing show promise and which ones would require significant algorithmic improvements to become practically-relevant. 

To establish reliable guidelines, or lower bounds for the required speedup of a quantum computer, we err on the side of being optimistic for quantum and overly pessimistic for classical computing. {Despite our overly-optimistic assumptions, our analysis will show that a wide range of often-cited applications is unlikely to result in a practical quantum advantage without \emph{significant} algorithmic improvements.} We compare the performance of only a single classical chip that is fabricated today  similar to the one used in the NVIDIA A100 GPU  which fits around 54 billion transistors~\cite{nvidia2020} with an optimistic assumption for a hypothetical quantum computer that may be available in the next decades with 10,000
error-corrected logical qubits, 10 $\mu s$ gate time for logical operations, the ability to simultaneously perform gate operations on all qubits and all-to-all connectivity for fault tolerant two-qubit gates.\footnote{Note that no quantum error correction scheme exists today that allows simultaneous execution of gates and all-to-all connectivity without at least a $O(\sqrt{N})$ slowdown for $N$ qubits.}

\begin{table}[h!]
\begin{center}
	\begin{tabular}{ l r r r}
		\  & \textbf{GPU} & \textbf{ASIC} & \textbf{Future Quantum} \\ 
		\hline
		\textbf{I/O bandwidth} & 10,000 Gbit/s & 10,000 Gbit/s & 1 Gbit/s \\  
		\hline
		\textbf{Operation throughput} & \ & \ & \\
		16-bit floating point & 195 Top/s & 550 Top/s & 10.5 kop/s \\
		32-bit integer & 9.75 Top/s & 215 Top/s & 0.83 kop/s\\
		binary (boolean logical) & 4,992 Top/s & 77,000 Top/s & 235 kop/s\\
		\hline
	\end{tabular}
\end{center}
\caption{\textbf{Performance comparison}. We compare the peak performance of a single classical chip that can be manufactured today (similar to an NVIDIA A100 GPU, or an ASIC with a similar number of transistors) with a future quantum computer with 10,000 error-corrected logical qubits, 10$\mu s$ gate time for logical operations and all-to-all connectivity. We consider an estimate of the I/O bandwidth (namely the number of operations per second) and three types of operations: logical binary operations, 16-bit floating point, 32-bit integer or fixed-point arithmetic multiply add operations.}
\end{table}

\paragraph{I/O bandwidth}  We first consider the fundamental I/O bottleneck that limits quantum computers in their interaction with the classical world, which determines bounds for data input and output bandwidths. Scalable implementations of quantum random access memory (QRAM~\cite{Giovannetti_2008,Giovannetti_20082}) demand a fault-tolerant error corrected implementation and the bandwidth is then fundamentally limited by the number of quantum gate operations or measurements that can be performed per unit time. {We assume only a single gate operation per input bit.} For our optimistic future quantum computer the resulting rate is 10,000 times smaller than for an existing classical chip (see Table 1). We immediately see that any problem that is limited by accessing classical data, such as search problems in databases, will be solved faster by classical computers. Similarly, a potentially exponential quantum speedup in linear algebra problems \cite{hhl}, vanishes when the matrix has to be loaded from classical data, or when the full solution vector should be read out. More generally, quantum computers will be practical for \emph{``big compute” problems on small data}, not big data problems.

\paragraph{Crossover scale}  With quantum speedup, asymptotically fewer operations will be needed on a quantum computer than on a classical computer. Due to the high operational complexity and slower gate operations, however, each operation on a quantum computer will be slower than a corresponding classical one. As sketched in Figure 1, classical computers will thus always be faster for small problems and quantum advantage is realized beyond a problem-dependent crossover scale where the gain due to quantum speedup overcomes the constant slowdown of the quantum computer. To have real practical impact, the crossover time needs to be short, not more than weeks. Constants matter in determining the utility for  applications, as with any runtime estimate in computing.

\paragraph{Compute performance}  To model performance, we employ the well-known work-depth model from classical parallel computing to determine upper bounds of classical silicon-based computations and an extension for quantum computations. In this model, the work is the total number of operations and applies to both classical and quantum executions. In Table 1 we provide concrete examples using three types of operations: logical operations, 16-bit floating point, and 32-bit integer or fixed-point arithmetic operations for numerical modeling.  For the quantum costs, we consider only the most expensive parts in our estimates, again benefiting quantum computers: For arithmetic, we count just the dominant cost of multiplications, assuming that additions are free. Furthermore, for floating point multiplication, we consider only the cost of the multiplication of the mantissa (10 bits in fp16). We ignore all further overheads incurred by the quantum algorithm due to reversible computations, as well as the significant cost of mapping to a specific hardware architecture with limited qubit connectivity.

\paragraph{Crossover times for classical and quantum computation}  To estimate lower bounds for the crossover times, we next consider that while both classical and quantum computers have to evaluate the same functions (usually called oracles) that describe a problem, quantum computers require fewer evaluations thereof due to quantum speedup. At the root of many quantum acceleration proposals lies a quadratic quantum speedup, including the well-known \emph{Grover algorithm}~\cite{grover1,grover2}. For such an algorithm, a problem that needs $X$ function calls on a quantum computer requires quadratically more, namely on the order of $X^2$ calls on a classical computer. To overcome the large constant performance difference between a quantum computer and a classical computer, which Table 1 shows to be more than a factor of $10^{10}$, a large number of function calls $X\gg 10^{10}$ is needed for the quantum speedup to deliver a practical advantage. In Table 2, we estimate upper bounds for the complexity of the function that will lead to a cross-over time of $10^6$ seconds, or roughly two weeks.

\begin{table}[h!]
	\begin{center}
		\begin{tabular}{ l R{3.8cm} R{3.8cm} R{3.8cm}}
			& \multicolumn{3}{c}{\textbf{Maximum number of operations for practical}}\\
			\textbf{Operation type}  & \textbf{quadratic speedup} & \textbf{cubic speedup} & \textbf{quartic speedup} \\ 
			\hline
			16-bit floating point & 0.2 & 45,800 & 2,800,000 \\
			32-bit integer & 0.003 & 1,630 & 130,000\\
			Binary (logical) & 68 & 12,500,000 & 712,000,000\\
			\hline
		\end{tabular}
	\end{center}
	\caption{\textbf{Crossover operation counts for quantum algorithms with quadratic, cubic, and quartic speedups}. We determine the number of operations that can be afforded per function call (see Figure 1) for a quantum computer to show an advantage over a classical computer using a quantum algorithm with quadratic, cubic, and quartic quantum speedup. {The number of oracle calls required to reach the crossover point with a quadratic, cubic, and quartic speedup is computed using the relative runtimes of a single oracle evaluation, and the total runtime of $10^6$ seconds is then used to compute how many basic operations can be afforded in each oracle call.} Since we make optimistic assumptions for a future quantum computer, we ignore overheads of reversible arithmetic for quantum computing and limit the classical computer to a single chip that can be manufactured today. The actual crossover operation counts will be significantly smaller. A similar analysis for quantum algorithms with exponential speedups yields promising operation budgets for all datatypes.}
\end{table}

We see that with quadratic speedup even a single floating point or integer operation leads to crossover times of several months. Furthermore, at most 68 binary logical operations can be afforded to stay within our desired crossover time of two weeks, which is too low for any non-trivial application. Keeping in mind that these estimates are pessimistic for classical computation (a single of today’s classical chips) and overly optimistic for quantum computing (only considering the multiplication of the mantissa and assuming all-to-all qubit connectivity), we come to the clear conclusion that quadratic speedups are insufficient for practical quantum advantage.  The numbers look better for cubic or quartic speedups where thousands or millions of operations may be feasible, and we hence conclude, similarly to Babbush et al.~\cite{Babbush2021}, that at least cubic or quartic speedups are required for a practical quantum advantage, and taking into account . 

{As a result of our overly-optimistic assumptions in favor of quantum computing,} these conclusions will remain valid even with significant advances in quantum technology of multiple orders of magnitude.

\paragraph{Practical and impractical applications}  We can now use the above considerations to discuss several classes of applications where our fundamental bounds draw a line for quantum practicality. The most likely problems to allow for a practical quantum advantage are those with exponential quantum speedup. This includes the simulation of quantum systems for problems in chemistry, materials science, and quantum physics, as well as cryptanalysis using Shor’s algorithm~\cite{shor}. The solution of linear systems of equations for highly structured problems~\cite{hhl} also has an exponential speedup, but the I/O limitations discussed above will limit the practicality and undo this advantage if the matrix {has to be loaded from memory instead if being computed based on limited data or} knowledge of the full solution is required (as opposed to just some limited information obtained by sampling the solution).

Equally importantly, we identify {likely} dead ends in the maze of applications. A large range of problem areas with quadratic quantum speedups, such as many current machine learning training approaches, accelerating drug design and protein folding with Grover's algorithm, speeding up Monte Carlo simulations through quantum walks, as well as more traditional scientific computing simulations including the solution of many non-linear systems of equations, such as fluid dynamics in the turbulent regime, weather, and climate simulations will not achieve quantum advantage with current quantum algorithms in the foreseeable future. We also conclude that the identified I/O limits constrain the performance of quantum computing for big data problems, unstructured linear systems, and database search based on Grover’s algorithm such that a speedup is unlikely in those cases. {Furthermore, Aaronson et al.~\cite{aaronson21} show that the achievable quantum speedup of unstructured black-box algorithms is limited to $\mathcal{O}(N^4)$. This  implies that any algorithm achieving higher speedup must exploit structure in the problem it solves. }

These considerations help with separating hype from practicality in the search for quantum applications and can guide algorithmic developments. Specifically, our analysis shows that 1) it is necessary for the community to focus on super-quadratic speedups, ideally exponential speedups and 2) one needs to carefully consider I/O bottlenecks when deriving algorithms to exploit quantum computation best. {Therefore, }\emph{the most promising candidates for quantum practicality are small-data problems with exponential speedup}. {Specific examples where this is the case are} quantum problems in chemistry and materials science {\cite{feynman1981simulating}}, which we identify as the most promising application.
We recommend to use precise requirements models~\cite{resest} to get more reliable and realistic (less optimistic) estimates in cases where our rough guidelines indicate a potential practical quantum advantage.

\section*{Methods}

Here we provide more details for how we obtained the numbers above. 
We compare our quantum computer with a single microprocessor chip similar to the one used in the NVIDIA A100 GPU~\cite{nvidia2020}. 
The A100 chip is around $850 mm^2$ in size and manufactured in TSMC's 7nm N7 silicon process.
A100 shows that such a chip fits around 54.2 billion transistors and can operator at a cycle time of around 0.7ns. 

\subsection*{Determining peak operation throughputs}
In Table~1, we provide concrete examples using three types of operations: logical operations, 16-bit floating point, and 32-bit integer arithmetic operations for numerical modeling. Other datatypes could be modeled using our methodology as well.

\paragraph{Classical NVIDIA A100}
According to its datasheet, NVIDIA’s A100 GPU, a SIMT-style von Neumann load store architecture, delivers 312 tera-operations per second (Top/s) with half precision floating point (fp16) through tensor cores and 78 Top/s through the normal processing pipeline. NVIDIA assumes a 50/50 mix of addition and multiplication operations and thus, we divide the number by two, yielding 195 Top/s fp16 performance. The datasheet states 19.5 Top/s for 32-bit integer operations, again assuming a 50/50 mix of addition and multiplication, leading to an effective 9.75 Top/s. The binary tensor core performance is listed as 4,992 Top/s with a limited set of instructions.

\paragraph{Classical Special Purpose ASIC}
Our main analysis assumes that we build a special-purpose ASIC using a similar technology. 
If we were to fill the equivalent chip-space of an A100 with a specialized circuit, we would use existing execution units, for which the size is typically measured in gate equivalents (GE). 
A 16-bit floating point unit (FPU) with addition and multiplication functions requires approximately 7 kGE, a 32-bit integer unit requires 18 kGE~\cite{mach2020}, and we assume 50 GE for a simple binary operation. 
All units include operand buffer registers and support a set of programmable instructions. 
We note that simple addition or multiplication circuits would be significantly cheaper. 
If we assume a transistor-to-gate ratio of 10~\cite{jones2018} and that 50\% of the total chip area is used for control logic of a dataflow ASIC with the required buffering, we can fit $54.2B/(7k\cdot 10\cdot 2)=387k$ fp16 units.
Similarly, we can fit $54.2B/(18k\cdot 10\cdot 2)=151k$ int32, or $54.2B/(50\cdot 10\cdot 2)=54.2M$ bin2 units on our hypothetical chip. 
Assuming a cycle time of 0.7ns, this leads to a total operation rate of 0.55 fp16, 0.22 int32, and 77.4 bin Pop/s for an application-specific ASIC with the A100's technology and budget. The ASIC thus leads to a raw speedup between roughly 2 and 15x over a programmable circuit. Thus, on classical silicon, the performance ranges roughly between $10^{13}$ and $10^{16}$ op/s for binary, int32, and fp16 types.

\paragraph{Hypothetical future quantum computer}
To determine the costs of N-bit multiplication on a quantum computer, we choose the controlled adder from Gidney~\cite{Gidney2018halvingcostof} and implement the multiplication using N single-bit controlled adders, each requiring $2N$ CCZ magic states. These states are produced in so called ``magic state factories'' that are implemented on the physical chip. 
While the resulting multiplier is entirely sequential, we found that this construction allows for more units to be placed on one chip than for a low-depth adder and/or for a tree-like reduction of partial products since (1) the number of CCZ states is lower (and thus fewer magic state factories are required) and (2) the number of work-qubits is lower. The resulting multiplier has a CCZ-depth and count of $2N^2$ using $5N-1$ qubits ($2N$ input, $2N-1$ output, $N$ ancilla for the addition). 

To compute the space overhead due to CCZ factories, we first use the analysis of Gidney and Fowler~\cite{Gidney2019efficientmagicstate} to compute the number of physical qubits per factory when aiming for circuits (programs) using $\approx10^{8}$ CCZ magic states with physical gate errors of $10^{-3}$. We approximate the overhead in terms of logical qubits by dividing the physical space overhead by $2d^2$, where we choose the error-correcting code distance $d=31$ to be the same as the distance used for the second level of distillation~\cite{Gidney2019efficientmagicstate}.
Thus we divide Gidney and Fowler's 147,904 physical qubits per factory (for details consult the ancillary spreadsheet (field B40) of Gidney and Fowler) by $2d^2=2\cdot 31^2$ and get an equivalent space of 77 logical qubits per factory. 
%

For the multiplier of the 10-bit mantissa of an fp16 floating point number, we need $2\cdot 10^2=200$ CCZ states and $5\cdot 10=50$ qubits. 
Since each factory takes 5.5 cycles~\cite{Gidney2019efficientmagicstate} and we can pipeline the production of CCZ states, we assume 5.5 factories per multiplication unit such that multipliers don't wait for magic state production on average. Thus, each multipler requires 200 cycles and $5N+5.5\cdot 77=50+5.5\cdot 77=473.5$ qubits. With a total of 10,000 logical qubits, we can implement $21$ 10-bit multipliers on our hypothetical quantum chip. With 10$\mu s$ cycle time, the 200 cycle latency, we get the final rate of less than $10^5 cycle/s / (200 cycle/op) \cdot 21 = 10.5k op/s$. For int32 (N=32), the calculation is equivalent. 
For binary, we assume two input and one output qubit for the (binary) adder (Toffoli gate) which does not need ancillas.
The final results are summarized in Table 1.

\subsection*{A note on parallelism}

{We assumed massively parallel execution of the oracle on both the classical and quantum computer (i.e., oracles with a depth of one). If the oracle does not admit such parallelization, e.g., if depth = work in the worst case scenario, then the comparison becomes more favorable towards the quantum computer. One could model this scenario by allowing the classical computer to only perform one operation per cycle. With a 2 GHz clock frequency, this would mean a slowdown of about 100,000 times for fp16 on the GPU. In this \emph{extremely unrealistic} algorithmic worst case, the oracle would still have to consist of only several thousands of fp16 operations with a quadratic speedup. However, we note that in practice, most oracles have low depth and parallelization across a single chip is achievable, which is what we assumed in the main text.}

\subsection*{Determining maximum operation counts per oracle call}
In Table 2, we list the maximum number of operations of a certain type that can be run to achieve a quantum speedup within a runtime of $10^{6}$ seconds (a little more than two weeks).
The maximum number of classical operations that can be performed with a single classical chip in $10^6$ seconds would be: 0.55 fp16, 0.22 int32, and 77.4 bin Zop. Similarly, assuming the rates from Table 1, for a quantum chip: 7, 4, 2,350 Gop, respectively.

We now assume that all calculations are used in oracle calls on the quantum computer and we ignore all further costs on the quantum machine. 
We start by modeling algorithms that provide polynomial $X^k$ speedup, for small constants $k$. For example, for Grover’s algorithms~\cite{grover1}, $k=2$. It is clear that quantum computers are asymptotically faster (in the number of oracle queries) for any $k>1$. However, we are interested to find the oracle complexity (i.e., the number of operations required to evaluate it) for which a quantum computer is faster than a classical computer within the time-window of $10^6$ seconds.

Let the number of operations required to evaluate a single oracle call be $M$ and let the number of required invocations be $N$. 
It takes a classical computer time $T_c = N^k\cdot M\cdot t_c$, whereas a quantum computer solves the same problem in time $T_q = N\cdot M\cdot t_q$, where $t_c$ and $t_q$ denote the time to evaluate an operation on a classical and on a quantum computer, respectively. 
By demanding that the quantum computer should solve the problem faster than the classical computer and within $10^6$ seconds, we find

$$\sqrt[k-1]{\frac{t_q}{t_c}}\leq N\leq \frac{10^6}{t_q\cdot M},$$

which allows us to compute the maximal number of basic operations per oracle evaluation such that the quantum computer still achieves a practical speedup:

$$M\leq 10^6\cdot \sqrt[k-1]{\frac{t_c}{t^k_q}}.$$

\subsection*{Determining I/O bandwidth}
We use the I/O bandwidth specified in NVIDIA's A100 datasheet for our classical chips. For the quantum computer, we assume that one quantum gate is required per bit of I/O. Using all 10,000 qubits for reading/writing, this yields an estimate of the I/O bandwidth $B\approx \frac{10,000}{10^{-5}}=1$ Gbit/s.

\subsection*{Acknowledgments}
We thank Luca Benini for helpful discussions about ASIC and processor design and related overheads and Wim van Dam and all anonymous reviewers for comments that improved an earlier draft of this work.

\bibliographystyle{plain} 
\bibliography{quantum_vs_classical} 

\end{document}